\documentclass[apj]{emulateapj}
\usepackage{amsmath}
\usepackage{graphicx}
\usepackage{epsfig}
\usepackage{float}
\usepackage{amssymb}
\usepackage{subfigure}
\usepackage{natbib}
\usepackage{color}
\usepackage[normalem]{ulem}  

\newcommand{\goo}{\,\raisebox{-.5ex}{$\stackrel{>}{\scriptstyle\sim}$}\,}

\shorttitle{Tabulated equation of state for supernova matter}
\shortauthors{Buyukcizmeci, Botvina and Mishustin}

\begin{document}

\title{Tabulated equation of state for supernova matter including full nuclear ensemble}

\author{N.~Buyukcizmeci$^{1,2}$, A.S.~Botvina$^{1,3,4}$, I.N.~Mishustin$^{1,5}$}

\affil{$^1$Frankfurt Institute for Advanced Studies, J.W. Goethe University, D-60438  Frankfurt am Main, Germany\\
$^2$Department of Physics, Selcuk University, 42079 Kampus, Konya, Turkey\\
$^3$Institute for Nuclear Research, Russian Academy of Sciences, 
117312 Moscow, Russia\\
$^4$Institut f\"{u}r Kernphysik, J. Gutenberg Universit\"{a}t, 55099 Mainz, Germany\\
$^5$Kurchatov Institute, Russian Research Center, 123182 Moscow, Russia.}

\begin{abstract}
This is an introduction to the tabulated data  base of stellar matter properties calculated within the framework of the Statistical Model for Supernova Matter (SMSM). The tables present thermodynamical characteristics and nuclear abundances for 31 values of baryon density (10$^{-8}<\rho/\rho_0<$0.32, $\rho_0$=0.15 fm$^{-3}$ is the normal nuclear matter density), 35 values of temperature ($0.2<T<25$ MeV) and 28 values of electron-to-baryon  ratio ($0.02<Y_e<0.56$). The properties of stellar matter in $\beta$-equilibrium are also considered. The main ingredients of the SMSM are briefly outlined, and the data structure and content of the tables are explained.
\end{abstract}

\journalinfo{Published in ApJ, ***, ** (201*)}

\maketitle

\section{Introduction}

As established in intensive experimental studies of last 20 years, many nuclear reactions lead to the formation of thermalized nuclear systems characterized by subnuclear densities and temperatures of 3-8 MeV. De-excitation of such systems goes through nuclear multifragmentation, i.e. break-up into many excited fragments and nucleons. As it is generally accepted, thermal and chemical equilibrium can be established in such multifragmentation reactions. Transport theoretical calculations of both central heavy-ion collisions around the Fermi-energy and peripheral heavy-ion collisions predict momentum distributions of nucleons which are similar to equilibrium ones after $\approx 100$ fm/c (see references in the topical issue of \citet{EPJA}). In early 80s several versions of the statistical approach were proposed to describe multifragmentation of highly-excited equilibrated sources, see e.g. \citet{Randrup81,Gross82,Bondorf85}.
These models were able to  describe many characteristics of nuclear fragments observed in nuclear  experiments: multiplicities of intermediate-mass fragments, charge and isotope distributions, event by event correlations of fragments (including fragments of different sizes), their angular and velocity correlations, and other observables (\citet{Gross,Bondorf,Souliotis,Ogul,Botvina90,ALADIN,EOS,MSU,Dag,INDRA,FASA,Hauger,Iglio, Hudan,Wang,Viola,Rodionov,Pienkowski}). The temperature and density of nuclear matter at the stage of fragment formation can also be established reliably in experiment by measuring relative velocities of fragments and ratios of isotope yields. 

\begin{figure} 
\begin{center}
\includegraphics[width=8cm,height=8cm]{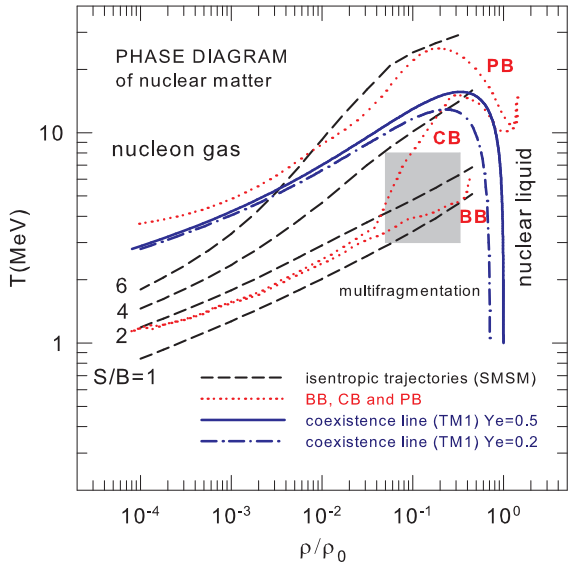} 
\end{center}
\caption{\label{fig_phase}\small{Nuclear phase diagram in the 'temperature -- baryon density' plane. Solid and dashed-dotted blue lines indicate boundaries of the liquid-gas coexistence region for symmetric ($Y_e=0.5$) and asymmetric matter ($Y_e=0.2$) calculated with TM1 interactions (\citet{Suga94}). The shaded area corresponds to typical conditions for nuclear multifragmentation reactions. The dashed black lines are isentropic trajectories characterized by constant entropy per baryon, $S/B=$1, 2, 4 and 6 calculated with SMSM (\citet{BotvinaMishustin:2010}). The dotted red lines show model results of \citet{Sumi} for BB (just before the bounce), CB (at the core bounce) and PB (the post bounce) in a core-collapse supernova. This figure is taken from \citet{nihal2013}(Color version online.). 
}}
\end{figure}

A nuclear matter of similar type is expected to be formed in astrophysical processes, such as collapse of massive stars and supernova explosions. To compare thermodynamical conditions obtained in nuclear reactions and in astrophysics, in Fig.~\ref{fig_phase} (\citet{nihal2013}) we show the phase diagram for symmetric and asymmetric nuclear matter for a corresponding range of densities and temperatures. Typical conditions associated with multifragmentation reactions are indicated by the shaded area in Fig.~\ref{fig_phase}. These reactions give us a chance to study hot nuclei in the environment of other nuclear species in thermodynamical equilibrium as, we expect in hot stellar matter at subnuclear densities. The properties of such nuclei can be directly extracted from experimental data and then this information can be used for more realistic calculations of nuclear composition in stellar matter. This is a new possibility to study stellar matter besides of theoretical approaches which use nuclear forces extracted from experimental study of nuclear structure. As one can see from Fig.~\ref{fig_phase}, in the course of massive star collapse the stellar nuclear matter passes exactly through the multifragmentation region with typical entropy per baryon $S/B=1-4$.
The electron fraction $Y_e$, which is equal to the total proton fraction, in the supernova core varies from 0.1 to 0.5. In the two-phase coexistence region (below the solid and dot-dashed lines) at densities $\rho \approx 0.3-0.8\rho_0$ the matter should be in a mixed phase, which is strongly inhomogeneous state with intermittent dense and dilute regions. In the coexistence region at lower densities, $\rho<0.3\rho_0$ the nuclear matter breaks up into compact nuclear droplets surrounded by nucleons. These relatively low densities dominate during the main stages of stellar collapse and explosion. Under such conditions one can expect a mixture of different species including nucleons, light and heavy nuclei. In Fig.~\ref{fig_phase} we demonstrate also isentropic trajectories of nuclear matter (dashed curves) as well as dynamical trajectories of density and temperature inside the supernova core (red dotted curves), taken from the supernova simulation of a 15 solar mass progenitor (\citet{Sumi}). The snapshots in the supernova dynamics are selected for three stages: before bounce (BB), at core bounce (CB), and after bounce (PB). At the gravitational collapse of the iron core just before bounce (BB), the density and temperature roughly follow the isentropic curves with $S/B=1-2$. At the core bounce (CB) the central density increases just above the nuclear matter density $\rho_0$=0.15 $fm^{-3}$. After that  the temperature of the inner core becomes higher than 10 MeV due to the passage of the shock wave. The temperature of the whole supernova core is still high at 150 ms after the bounce (PB), when the shock is stalled around 130 km. This 1D simulation, does not lead to a successful explosion. 

One can see that thermodynamic conditions inside the supernova core cover interesting regions of the phase diagram. The BB and CB trajectories pass through the multifragmentation region, which is explored in heavy-ion collisions at intermediate energies. The trajectories CB and PB traverse the phase boundary between the mixture of nuclei and the nuclear gas. This region is dominated by nucleons and light nuclei (n, p, d, t, $^3$He,$^4$He). It is well known, see e.g. Ref. \citet{Arcones:etal:2008}, that dynamics of the shock wave is strongly affected by the neutrino-induced reactions on nucleons and nuclei. Therefore, it is very important to determine the realistic composition of hot stellar matter in this region. This can be done only by considering the full ensemble of nuclear species without artificial constraints (\citet{Japan,BotvinaMishustin:2004}). The nuclear composition and thermodynamic properties of nuclear matter under supernova conditions were studied recently within different approaches  (\citet{BotvinaMishustin:2010,HempelSchaffnerBielich:2010,Hempel12,Typel:etal:2010,Hempel11b,SumiyoshiRopke:2008,nihal2013}.

\section{Statistical Model for Supernova Matter (SMSM)}

The Statistical Model for Supernova Matter (SMSM) was developed in Refs.~(\citet{BotvinaMishustin:2004,BotvinaMishustin:2010}) as a direct generalization of the Statistical Multifragmentation Model -- SMM (\citet{Bondorf}). The SMM was successfully used for description of nuclear multifragmentation reactions (\citet{Ogul,ALADIN,EOS,MSU,Dag,INDRA,FASA,Hauger,Iglio, Hudan,Wang,Viola}). This gives us confidence that this model can realize a realistic approach to  clustered nuclear matter under astrophysical conditions, as discussed in Introduction. We treat supernova matter as a mixture of nuclear species, electrons, and photons in statistical equilibrium. Below we briefly present the model with latest modifications.

\subsection{Equilibrium conditions}

For describing stellar matter we adopt Grand Canonical approximation where macroscopic states are characterized by temperature $T$ and chemical potentials of the constituents. 
The chemical potentials for nuclear species $(A,Z)$ are expressed as
\begin{equation} \label{chem}
 \begin{array}{ll}
\mu_{AZ}=A\mu_B+Z\mu_Q~,
 \end{array}
\end{equation}
For protons $\mu_{\rm p}=\mu_{\rm B}+ \mu_{\rm Q}$, for neutrons $\mu_{\rm n} = \mu_{\rm B}$. Two chemical potentials  $\mu_B$ and $\mu_Q$ are found from the conservation laws for baryon number $B$ and electric charge $Q$ in the normalization volume $V$:
\begin{eqnarray}  \label{conserv-laws}
\rho=\frac{B}{V}=\sum_{AZ}A\rho_{AZ}~,~\nonumber\\
\rho_Q=\frac{Q}{V}=\sum_{AZ}Z\rho_{AZ}-\rho_e=0~.
\end{eqnarray}
Here, $\rho_{AZ}$ is the average density of nuclear species with mass $A$ and charge $Z$ (see below), $\rho_e=\rho_{e^-}-\rho_{e^+}$ is the net electron density. The second equation requires any macroscopic volume of the star ($V$) to be electrically neutral. 

The lepton number conservation is a valid concept only if $\nu$ and $\tilde{\nu}$ are trapped in the matter within the neutrino-sphere (\citet{Prakash}). If they escape freely from the star, the lepton number conservation is irrelevant and $\mu_L=0$. In this case two remaining chemical potentials are determined by the equations (\ref{conserv-laws}). In the $\beta$-equilibrated matter the electron chemical potential $\mu_e$ is found from the condition $\mu_e=\mu_n-\mu_p$. However, $\beta$-equilibrium may not be achieved in a fast explosive process. In this case it is more appropriate to perform calculations for fixed values of electron fraction $Y_e$. Then $\mu_e$  is determined from the given electron density  $\rho_e=Y_e \rho$.

\subsection{Ensemble of nuclear species} 

The nuclear component of stellar matter is represented as a mixture of gases of different species (A,Z) including nuclei and nucleons. It is convenient to introduce the numbers of particles of different kind $N_{AZ}$ in a normalization volume $V$. In SMSM we use the Grand Canonical version of the SMM formulated in Ref. (\citet{Botvina85}), and developed further in 
Ref. (\citet{traut}). The corresponding thermodynamic potential of the system can be expressed as 
\begin{eqnarray} 
\Omega(T,\mu_B,\mu_Q, \{N_{AZ}\},V)= \nonumber \\
F^{\rm tr}(T,\{N_{AZ}\},V_f)+\sum_{AZ}N_{AZ}F_{AZ}(T,\rho) \nonumber \\
-\mu_B\sum_{AZ}AN_{AZ}-\mu_Q\sum_{AZ}ZN_{AZ}~, \label{Omega}
\end{eqnarray} 
where the first term accounts for the translational degrees of freedom of nuclear fragments and the second term is associated with their internal degrees of freedom. Assuming the Maxwell-Boltzmann statistics for all nuclear species, the translational free energy can be explicitly written as\footnote{Here and below we use units with $\hbar=c=1$.}
\begin{equation} \label{Ftr}
F^{\rm tr}(T,\{N_{AZ}\},V)=-T\sum_{AZ}N_{AZ}\left[\ln{\left(\frac{g_{AZ}V_fA^{3/2}}{N_{AZ}\lambda_T^3}\right)}+1\right]~,
\end{equation}
where $g_{AZ}$ is the ground-state degeneracy factor for species $(A,Z)$, $\lambda_T=\left(2\pi/m_NT\right)^{1/2}$ is the nucleon thermal wavelength, $m_N \approx 939$ MeV is the average nucleon mass and $V_f$ is so called free volume of the system, which accounts for the finite size of nuclear species and is only a fraction of the total volume $V$. We assume that all nuclei with $A>4$ have normal nuclear density $\rho_0\approx 0.15$ fm$^{-3}$, so that the proper volume of a nucleus with mass $A$ is $A/\rho_0$. At relatively low densities considered here one can adopt the excluded volume approximation, $V_f/V \approx \left(1-\rho/\rho_0\right)$. This approximation is commonly accepted in statistical models, see extended discussion of this problem in Ref. \citet{Sagun}. 
Certain information about the free volume in multifragmentation reactions has been extracted from analysis of experimental data  (\citet{EOS}). 

\subsection{The internal free energy of fragments}

The internal excitations of nuclei play an important role in regulating their abundances, since they increase significantly their entropy. Some authors (see, e.g., Ref.~(\citet{Japan})) limit the excitation spectrum by particle-stable levels known for low excited nuclei. Within the SMM we follow quite different philosophy motivated by experimental investigations of nuclear disintegration reactions. Namely, we assume that excited states are populated according the internal temperature of nuclei, which is assumed to be the same as the temperature of surrounding medium. In this case not only particle-stable states but also particle-unstable states will contribute to the excitation energy and entropy. This assumption can be justified by the dynamical equilibrium between emission and absorption processes in the hot medium. Moreover, in the supernova environment both the excited states and the binding energies of nuclei may be strongly affected by the surrounding matter. By this reason, we find it more appropriate to use an approach which can easily be generalized to include in-medium modifications. Namely, the internal free energy of species $(A,Z)$ with $A>4$ is parametrized in the spirit of the liquid drop model, which has been proved to be very successful in nuclear physics:
\begin{equation} \label{Fint}
F_{AZ}(T,\rho)=F_{AZ}^B+F_{AZ}^S+F_{AZ}^{\rm sym}+F_{AZ}^C~.
\end{equation}
Here the right hand side contains, respectively, the bulk, the surface, the symmetry and the Coulomb terms. The first three terms are taken in the standard form ( \citet{Bondorf}),
\begin{eqnarray}
F_{AZ}^B(T)=\left(-w_0-\frac{T^2}{\varepsilon_0}\right)A~, \\
F_{AZ}^S(T)=\beta_0\left(\frac{T_c^2-T^2}{T_c^2+T^2}\right)^{5/4}A^{2/3}~,\\
\label{fres}
F_{AZ}^{\rm sym}=\gamma \frac{(A-2Z)^2}{A}~,
\end{eqnarray}
where $w_0=16$ MeV, $\varepsilon_0=16$ MeV, $\beta_0=18$ MeV, $T_c=18$ MeV and $\gamma=25$ MeV are the model parameters which are extracted from nuclear phenomenology and provide a good description of multifragmentation data (\citet{Bondorf,ALADIN,EOS,MSU,INDRA,FASA,Dag}). However, these parameters, especially the symmetry coefficient $\gamma$, may be different in hot nuclei, and therefore they should be determined from corresponding experimental data, see discussions in Refs. (\citet{Botvina06,bulk,traut,Ogul}). 

In the electrically-neutral environment the nuclear Coulomb term should be modified to include the screening effect of electrons. This can be done, e.g., within the Wigner-Seitz approximation as was proposed in Refs.~(\citet{Lamb:1981,Lattimer:etal:1985}). One should imagine that the whole system is divided into spherical cells each containing one nucleus. The radius of the cell is determined by the condition that it contains the same number of electrons as the number of protons in the nucleus. The interaction between the cells is neglected. Then, assuming a constant electron density one obtains 
\begin{eqnarray} \label{fazc}
F_{AZ}^C(\rho)=\frac{3}{5}c(\rho)\frac{(eZ)^2}{r_0A^{1/3}}\cdot~~\\
c(\rho)=\left[1-\frac{3}{2}\left(\frac{\rho_e}{\rho_{0p}}\right)^{1/3}
+\frac{1}{2}\left(\frac{\rho_e}{\rho_{0p}}\right)\right]~,\nonumber
\end{eqnarray}
where $r_0=1.17$ fm, $\rho_e=Y_e\rho$ is the average electron density and $\rho_{0p}=(Z/A)\rho_0$ is the proton density inside the nuclei. The screening function $c(\rho)$ is 1 at $\rho_e=0$ and 0 at $\rho_e=\rho_{0p}$. In fact, in this work all presented results are obtained with the approximation, $\rho_e/\rho_{0p} \approx \rho/\rho_0$, as in Ref.~(\citet{Lattimer:etal:1985}), which works well when neutrons are mostly bound in nuclei, so that $\rho_{0p}\approx Y_e\rho_0$. Although this approximation gives satisfactory results 
in many cases, it may deviate at very high temperatures and 
very low baryon densities. However, under these conditions the nuclei are getting smaller and the Coulomb interaction effects become less important. 
Generally, the reduction of the Coulomb energy due to electron screening favors the formation of heavy nuclei. 

In the SMM nucleons and light nuclei $(A \leq 4)$ are considered as structure-less particles characterized only by exact masses, proper volumes and spin degeneracy factors $g_{AZ}$ (\citet{Bondorf}): for nucleons $g_n=g_p=2$; for deuterons $g_{21}=3$; for $^3$H  $g_{31}=2$; for $^3$He $g_{32}=2$; for $^4$He $g_{AZ}=1$. Their Coulomb interaction is taken into account within the same Wigner-Seitz approximation. For all nuclear species with $A>4$ we use $g_{AZ}$=1, but include internal excitations.

\subsection{Thermodynamical quantities}

Within the Grand Canonical approximation the conservation laws Eqs. (\ref{conserv-laws}) are fulfilled only for the mean densities of nuclear species $\rho_{AZ}=\langle N_{AZ}\rangle/V$. The mean fragment numbers $\langle N_{AZ}\rangle$ in volume $V$ are found by minimizing the thermodynamic potential Eq.~(\ref{Omega}) with respect to fragment multiplicities $\{N_{AZ}\}$ at fixed $T$ and $V$. In the lowest-order approximation this gives
\begin{equation} \label{NAZ}
\langle N_{AZ}\rangle=g_{AZ}\frac{V_f}{\lambda_T^3}A^{3/2}
{\rm exp}\left[-\frac{1}{T}\left(F_{AZ}-\mu_{AZ}\right)\right]~,
\end{equation}
where $\mu_{AZ}$ is defined by Eq.~(\ref{chem}). The corrections to this expression become significant at higher densities and low temperatures when heavy nuclei are abundant.

The pressure associated with nuclear species can be calculated by differentiating Eq.~(\ref{Omega}) with respect to $V$ at fixed $T$ and $\{N_{AZ}\}$ that gives 
\begin{equation} \label{totnuc}
P_{\rm nuc}=P_{\rm tr}+P_{\rm C}~, 
\end{equation}
where first term comes from the translational motion of fragments and second term, from the density-dependent Coulomb interaction. Their explicit expressions are
\begin{equation} \label{pressure}
P_{\rm tr}=T\sum_{AZ}\rho_{AZ},~~~P_{\rm C}=\rho \sum_{AZ}\rho_{AZ} \frac{\partial F_{AZ}^C}{\partial \rho}~.
\end{equation}
As one can see from Eq.~(\ref{fazc}), the Coulomb term gives  a negative contribution to pressure (\citet{BPS}).

\begin{figure} 
\begin{center}
\includegraphics[width=8cm,height=9cm]{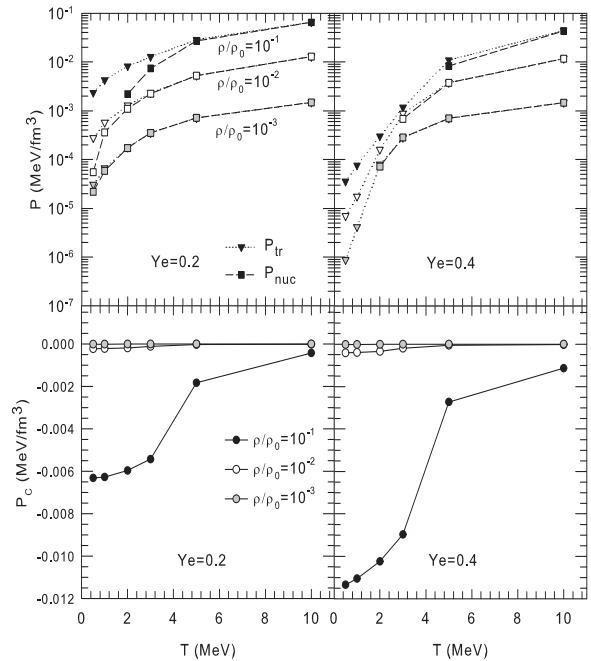}
\end{center}
\caption{\label{fig_pressure}\small{Comparison of results for the Coulomb pressure $P_{C}$ (bottom panels), 
translational pressure $P_{tr}$ and total nuclear pressure $P_{nuc}$ (top panels) as a function of temperature for $Y_e=0.2$, 0.4 and $\rho/\rho_0=10^{-1}-10^{-3}$.
 }}
\end{figure} 

In Fig.~\ref{fig_pressure}, we present the translational pressure $P_{\rm tr}$, Eq.~(\ref{pressure}), caused by single nucleons and nuclei. The total nuclear pressure $P_{\rm nuc}$ is also shown in Fig.~\ref{fig_pressure} as a function of temperature. It is important that at $T \geqslant 5$ MeV, when matter nearly completely dissociates into nucleons and lightest clusters, $P_C$ is close to zero. In this region the total nuclear pressure coincides with the pure nuclear pressure. The Coulomb pressure becomes very important when heavy clusters dominate in the system. One can see in Fig.~\ref{fig_pressure} that at low temperatures and high density the total nuclear pressure may be negative. In this case the nuclear clusterization is favorable for the collapse. However, the positive pressure of the relativistic electron Fermi-gas is considerably (more than an order of magnitude) larger. Therefore the total pressure $P_{\rm tot}$ given by the sum of the total nuclear, electron and photon pressures in such environment will always be positive and the condition of thermodynamical stability ( $\partial P_{\rm tot}/\partial \rho\geq 0$) will be fulfilled. 

Finally, the entropy of nuclear species is calculated by differentiating Eq.~(\ref{Omega}) with respect to $T$ at fixed $V$ and $\{N_{AZ}\}$ that gives

\begin{equation} \label{entropy}
S_{\rm nuc}=\sum_{AZ}N_{AZ}\left[\ln{\left(\frac{g_{AZ}V_fA^{3/2}}{N_{AZ}\lambda_T^3}\right)}+\frac{5}{2}\right]-\sum_{AZ}N_{AZ}\frac{\partial F_{AZ}}{\partial T}.
\end{equation}
Here the first term comes from the translational motion and the second one from internal excitations of fragments.

As follows from Eq.~(\ref{NAZ}), the fate of heavy nuclei depends strongly on the relationship between $F_{AZ}$ and $\mu_{AZ}$. In order to avoid an exponentially divergent contribution to the baryon density, at least in the thermodynamic limit ($A \rightarrow \infty$), inequality  $F_{AZ}\geqslant\mu_{AZ}$ must hold. The equality sign here corresponds to the situation when a big (infinite) nuclear fragment coexists with the gas of smaller clusters (\citet{Bugaev}). When $F_{AZ}>\mu_{AZ}$, only small clusters with nearly exponential mass spectrum are present. However, there exists a region of thermodynamic quantities corresponding to $F_{AZ}\approx\mu_{AZ}$ when the mass distribution of nuclear species is close to a power-law $A^{-\tau}$ with $\tau \approx 2$. This is a characteristic feature of the liquid-gas phase transition. The advantage of our approach is that we consider all the fragments present in this transition region, and, therefore, can study this phase of nuclear matter in all details. 

\subsection{Leptons and photons}

We assume that besides nuclear species the supernova matter also contains electrons, positrons and photons.\footnote{As has been already mentioned in Sect. 2.1, here we do not consider the situation when neutrinos also participate in statistical equilibrium.} At $T,\mu_e > m_e$ the pressure of the relativistic electron-positron gas can be written as
\begin{eqnarray} \label{Pel}
P_{e}=\frac{g_e\mu_e^4}{24\pi^2} \cdot ~~~~~~\\ \nonumber
\left[1+2\left(\frac{\pi T}{\mu_e}\right)^2+
\frac{7}{15}\left(\frac{\pi T}{\mu_e}\right)^4-\frac{m_e^2}{\mu_e^2}
\left(3+\left(\frac{\pi T}{\mu_e}\right)^2\right)\right] ,
\end{eqnarray}
where first-order corrections ($\sim m_e^2$) due to the finite electron mass is included, $g_e$=2 is the spin degeneracy factor for electrons. Due to the correction terms of order $(m_e/\pi T)^2$ and $(m_e/\mu_e)^2$ these
expressions can be used even at T, $\mu_e$ of order $m_e$. At these conditions the contributions
of electrons to the pressure and entropy become negligible. The corresponding expressions for net number density $\rho_e$ and entropy density $s_e$ are obtained from standard thermodynamic relations, $\rho_e=\partial P_e/\partial \mu_e$, and $s_e=\partial P_e/\partial T$, which give 
\begin{eqnarray} \label{eden}
\rho_e=\frac{g_e\mu_e^3}{6\pi^2}\left[1+\frac{1}{\mu_e^2}\left(\pi^2T^2-
\frac{3}{2}m_e^2\right)\right]~,\\
s_e=
\frac{g_e T\mu_e^2}{6}\left[1+\frac{7}{15}\left(\frac{\pi T}{\mu_e}\right)^2-
\frac{m_e^2}{2\mu_e^2}\right]~. 
\end{eqnarray}
The photons are always close to the thermal equilibrium, and they are treated as massless Bose gas with zero chemical potential. The corresponding density $\rho_{\gamma}$, energy density $e_{\gamma}$, pressure $P_{\gamma}$, and entropy density $s_{\gamma}$ of photons are given by standard formulae: 
\begin{equation} \label{gamma}
\rho_{\gamma}=\frac{g_{\gamma}\xi(3) T^3}{\pi^2 }~,~
e_{\gamma}=\frac{g_{\gamma}\pi^2 T^4}{30 }~,~
P_{\gamma}=\frac{e_{\gamma}}{3}~,~
s_{\gamma}=\frac{4 e_{\gamma}}{3 T},
\end{equation}
where $g_{\gamma}$=2. A possible participation in the statistical ensemble of $\nu_e, \tilde{\nu_{\rm e}}$, $\mu$, $\nu_{\rm \mu}$, and $\tilde{\nu_{\rm \mu}}$ is not considered in this work, so the current SMSM tables do not include these particles. However, the computer SMSM code includes the options of 
$\beta$-equilibrium, as well as the full lepton (including 
neutrino) conservation, therefore, such calculations are 
possible (see, \citet{BotvinaMishustin:2004,BotvinaMishustin:2010}).

The SMSM was realized with a generalization of the special 
computer code developed previously for the grand-canonical 
calculations of the full ensemble of nuclear species, as 
performed in Ref.~(\citet{traut}). 
All kinds of particles 
contribute to the free energy, pressure and other thermodynamical 
characteristics of the system, and we sum up all these contributions. 
The densities of all particles are calculated self-consistently by taking 
into account the relations between their chemical potentials. In the 
solution of the coupled equations we used the step-by-step approximation 
method and control the found potentials with the relative precision of 0.001.
We have checked that it is sufficient for our purposes. In this procedure 
we directly observe how the succesive steps approaching the correct values during the simulations. We take baryon number $B=$1000 and perform calculations for all fragments with 1$\leq A \leq$1000 and 0$\leq Z \leq A$ in the ensemble. This restriction on the size of 
nuclear fragments is fully justified in our case, since 
fragments with larger masses ($A>$1000) can be produced only at
very high densities $\rho \goo 0.3\rho_0$ 
(\citet{Lamb:1981,Lattimer:etal:1985}), which are 
appropriate for the regions deep inside the protoneutron star. 
Such nuclei, as well as the "pasta" phase at the high densities, are not considered here.

\section{The SMSM EOS Tables}

We have constructed the SMSM EOS tables which are available on the Web at \textit{ http://fias.uni-frankfurt.de/physics/mishus/research/smsm/}. Below we give the detailed information about the format of these tables and their possible applications.

\subsection{Description of the SMSM EOS Tables}

The SMSM EOS tables cover the following ranges of control parameters:
\begin{itemize}
\item {{\bf Temperature:} $T=0.2-25$ MeV; for 35 $T$ values.}
\item {{\bf Electron fraction $Y_e$:} $0.02-0.56$; linear mesh of $Y_e = 0.02$, giving 28 $Y_e$ values. It is equal to the total proton fraction $X_p$, due to charge neutrality.}
\item {{\bf Baryon number density fraction} $\rho/\rho_0=(10^{-8}-0.32)$, giving 31 $\rho/\rho_0$ values.}
\end{itemize}
In our calculations we consider all nuclear species with $1\leq A\leq 1000$ and $0\leq Z\leq A$. These restrictions on the fragment size are justified within the above-defined intervals of control parameters.  
The different $T$ (35 points) , $\rho/\rho_0$ (31 point) and $Y_e$ (28 points) values are listed in Table 1. They sum up to 30380 different grid points. We believe that physical conditions for the SMSM EOS are sufficiently well represented by this grid. Running time for calculations was approximately 7.8 days (Work station:4xDual core AMD 2.2 GHz processor). We have obtained the file \textit{SMSM-EOS-Tables.txt} (size 6.8 MB), containing 31497 lines.

We can certainly say that our model is not applicable at densities above
$0.3\rho_0$ and temperatures above 25 MeV. At higher densities the matter will be
rearranged into more complicated geometrical structures known as "pasta" phases (\citet{Lamb:1981,Newton}). We are not considering them in our present work. The assumption of statistical equilibrium requires also that the temperature is high enough to allow nuclear transformations leading to this equilibrium. As is commonly believed this could
be a good approximation at $T>1$ MeV. Nevertheless, for completeness we extend our
calculations to $T=0.2$ MeV, when the statistical equilibrium could be problematic.
\begin{table} [t]
\begin{center}
\begin{tiny}
\caption{Temperature, electron fraction and density fraction values of SMSM EOS tables.}
\centering
\begin{tabular}{cccccccc}
\hline 
& & & T(MeV)& & && \\
\hline
0.20 & 0.40 & 0.60 & 0.80 & 1.00 & 1.25 & 1.50 &\\
1.75 & 2.00 & 2.25 & 2.50 & 2.75 & 3.00 & 3.25 &\\
3.50 & 3.75 & 4.00 & 4.25 & 4.50 & 4.75 & 5.00 &\\
5.25 & 5.50 & 5.75 & 6.00 & 6.50 & 7.00 & 7.50& \\
8.00 & 9.00 & 10.00 & 12.00 & 15.00 & 20.00 & 25.00&\\
\hline
& & &$Y_e$& & & &\\
\hline 
0.02 & 0.04 & 0.06 & 0.08 & 0.10 & 0.12 & 0.14& \\
0.16 & 0.18 & 0.20 & 0.22 & 0.24 & 0.26 & 0.28 &\\
0.30 & 0.32 & 0.34 & 0.36 & 0.38 & 0.40 & 0.42 &\\
0.44 & 0.46 & 0.48 & 0.50 & 0.52 & 0.54 & 0.56 &\\
\hline
& & &$\rho / \rho_0$& & &&\\
\hline 
1E-8& 1.78E-8& 3.16E-8& 5.62E-8& 1E-7& 1.78E-7& 3.16E-7\\
5.62E-7& 1E-6& 1.78E-6& 3.16E-6& 5.62E-6& 1E-5& 1.78E-5\\
3.16E-5 & 5.62E-5& 1E-4& 1.78E-4& 3.16E-4& 5.62E-4& 1E-3\\
1.78E-3& 3.16E-3 & 5.62E-3& 1E-2& 1.78E-2& 3.16E-2& 5.62E-2\\
&&1E-1& 1.78E-1& 3.16E-1 &&\\
\hline
\end{tabular}
\end{tiny}
\end{center}
\end{table}


The information is stored in a format which is very similar to the tables of \citet{Shen} or \citet{HempelSchaffnerBielich:2010} or \citet{furusawa11} so that it can easily be implemented in running codes. The SMSM EOS tables file under the name \textit{"SMSM-EOS-Tables.txt"} is the main file which is available in the online journal.

In the beginning of file we write the general parameters of liquid-drop description: $w_0=16$ MeV, $\beta_0=18$ MeV, $T_c=18$ MeV, $\gamma=25$ MeV, $\varepsilon_0=16$ MeV, $r_0=1.17$ fm. The tables are written in the order of increasing $T$. In the tables, the values of $T$ (MeV) are given at the beginning of each block, and the blocks with different T are divided by the string of characters $cccccc$. For each $T$, we present the results in the order of increasing $Y_e$ and $\rho/\rho_0$ (see Table 1). We put also space lines between $Y_e$ blocks. The first three lines in SMSM EOS Tables are shown in Table 2 for guidance regarding its form and content. Each line of the table corresponds to one density grid point. It contain 20 different thermodynamic quantities which are defined as follows:
\begin{enumerate}
\item {$Y_e$, electron fraction:~$Y_e = \rho_e/\rho$,} 
\item {$\rho/\rho_0$, density fraction,}
\item {$X_n$, the mass fraction of free neutrons is given by:~$X_n = \rho_n/\rho$,}
\item {$X_p$, the mass fraction of free protons:~$X_p = \rho_p/\rho$,}
\item {$X_d$, the mass fraction of deuterons ($A = 2$, $Z = 1$) is given by:~$X_d = 2\rho_d/\rho$,}
\item {$X_t$, the mass fraction of tritons ($A=3$, $Z=1$) is given by:~$X_t = 3\rho_t/\rho$,}
\item {$X_{{^3}He}$, the mass fraction of helions ($A=3$, $Z=2$) is given by:~$X_{{^3}He} = 3 \rho_{{^3}He}/\rho$,}
\item {$X_{alpha}$, the mass fraction of alphas ($A=4$, $Z=2$) is given by:~$X_{alpha} = 4\rho_{alpha}/\rho$,}
\item {$X_{heavy}$, the mass fraction of heavy nuclei $(A>4, Z>2)$ is defined by:
\begin{eqnarray}
X_{heavy} = \frac{\sum_{A>4,Z>2}A\rho_{AZ}}{\rho},\nonumber
\end{eqnarray}
}
\item {$<A_{heavy}>$, the average mass number of heavy nuclei $(A>4, Z>2)$ is defined by:
\begin{eqnarray}
<A_{heavy}>=\frac{\sum_{A>4,Z>2}A\rho_{AZ}}{\sum_{A>4, Z>2}\rho_{AZ}},\nonumber
\end{eqnarray}
$<A_{heavy}>$ is set to zero if $X_{heavy}=0$.}
\item {$<Z_{heavy}>$, the average charge number of heavy nuclei ($A>4$, $Z>2$) is defined by:
\begin{eqnarray}
<Z_{heavy}>=\frac{\sum_{A>4,Z>2}Z\rho_{AZ}}{\sum_{A>4,Z>2}\rho_{AZ}},\nonumber
\end{eqnarray}
}
\item {$E_{\rm tot}$, total energy can be written as the sum of nuclear $E_{\rm nuc}$, electron $E_{\rm e}$, and photon $E_{\rm \gamma}$ energy contributions (MeV/nucleon).
\begin{eqnarray}
E_{\rm tot}=E_{\rm nuc}+E_{\rm e}+E_{\rm \gamma}.\nonumber
\end{eqnarray}}
\item {$E_{\rm nuc}$, the energy of nuclear species (MeV/nucleon), is calculated as $F+TS$, using Eqs. (\ref{Ftr}), 
(\ref{Fint}) and (\ref{entropy}).} 
\item {$S_{\rm tot}$, total entropy is the sum of nuclear $S_{\rm nuc}$, electron $S_{\rm e}$, and photon $S_{\rm \gamma}$ entropy contributions (1/nucleon),
\begin{equation}
S_{\rm tot}=S_{\rm nuc}+S_{\rm e}+S_{\rm \gamma}.\nonumber
\end{equation}}
\item {$S_{\rm nuc}$, nuclear entropy (1/nucleon) is given by Eq.~(\ref{entropy}).}
\item {$P_{\rm tot}$, total pressure is the sum of the pressure of nuclear species $P_{\rm nuc}$ (Eq.~(\ref{totnuc})), electrons $P_{\rm e}$ (Eq.~(\ref{Pel})), and photons $P_{\rm \gamma}$ (Eq.~(\ref{gamma})), (MeV/fm$^{3}$),
\begin{eqnarray} \label{Ptot}
P_{\rm tot}=P_{\rm nuc}+P_{\rm e}+P_{\rm \gamma}.\nonumber
\end{eqnarray}} 
\item {$P_{\rm {nuc}}$, total nuclear pressure is the sum of translational pressure and Coulomb pressure 
(Eq. (\ref{pressure})), (MeV/fm$^{3}$), }
\item {$\mu_e$, chemical potential of electrons, (MeV),}
\item {$\mu_B $, chemical potential of baryons, (MeV),}
\item {$\mu_Q$, chemical potential of charged particles, (MeV).}
\end{enumerate}

The thermodynamic consistency of SMSM EOS Table is guarantied by using the grand canonical ensemble with the thermodynamical potential (\ref{Omega}). This approximation should be good for large volumes of matter considered here. The numerical accuracy of our calculations is typically better than 1 \%. The mass conservation and electrical neutrality conditions are especially checked. For example, the mass fractions of the different particle species sum up to unity:
\begin{eqnarray}
\delta_X =1-X_n+X_p+X_d+X_t+X_{^3He}+X_{alpha}+X_{heavy}.\nonumber
\end{eqnarray}
$\delta_X$ is found around 0.1 per cent.

\begin{table} [t]
\begin{center}
\begin{tiny}
\caption{SMSM EOS Tables are available in a machine-readable form in the online journal. 
}
\begin{tabular}{ccccccc}
\hline
Column1 & Column2 & Column3 & Column4 & Column5 \\
$Y_{e}$ & $\rho/\rho_0$ & $X_{n}$ & $X_{p}$ & $X_d$\\
\hline
0.02 & 0.100E-07 & 0.9447 & 0.3514E-38 & 0.1248E-35 \\
0.02 & 0.178E-07 & 0.9445 & 0.7178E-39 & 0.4503E-36 \\
0.02 & 0.316E-07 & 0.9442 & 0.1388E-39 & 0.1609E-36 \\
\hline
Column6 & Column7 & Column8 & Column9 & Column10\\
$X_t$ & $X_{{^3}He}$ & $X_{alpha}$ & $X_{heavy}$ & $<A_{heavy}>$ \\
\hline
 0.2067E-26 & 0 & 0.2871E-24 & 0.05527 & 89.66\\
 0.1327E-26 & 0 & 0.6668E-25 & 0.05551 & 90.70\\
 0.8426E-27 & 0 & 0.1519E-25 & 0.05582 & 91.76\\
\hline
Column11 & Column12 & Column13 & Column14 & Column15\\
$<Z_{heavy}>$ & $E_{tot}$ & $E_{nuc}$ & $S_{tot}$ & S$_{nuc}$ \\ 
 &(MeV/nucleon)&(MeV/nucleon)&(1/nucleon)&(1/nucleon)\\
\hline
 33.44 & 0.08469 &-0.1659 & 13.81 & 12.13\\
 33.66 &-0.02678 &-0.1664 & 12.53 & 11.58\\
 33.88 &-0.09010 &-0.1675 & 11.57 & 11.03\\
\hline
Column16 & Column17 & Column18 & Column19 & Column20 \\
$P_{tot}$ & $P_{nuc}$ & $\mu_e$ & $\mu_B$ & $\mu_Q$ \\
(MeVfm$^{-3})$ &(MeVfm$^{-3}$) & (MeV) & (MeV) & (MeV)\\
\hline
 0.4077E-09 & 0.2816E-09 & 0.01716 &-2.063 &-16.96\\
 0.6277E-09 & 0.5011E-09 & 0.03048 &-1.948 &-17.28\\
 0.1019E-08 & 0.8906E-09 & 0.05394 &-1.833 &-17.60\\
\hline
\end{tabular}
\end{tiny}
\end{center}
\end{table}
\subsection{Application of the EOS tables and illustrative results}

In addition to standard output, i.e. the 20 different thermodynamic quantities listed above, like the pressure, entropy, energy, fractions of light particles and heavy nuclei (actually in \citet{Shen} case a single heavy nucleus), we provide a possibility to determine the ensemble-averaged densities of all heavy nuclei,$\rho_{AZ}$. This is achieved by first finding the chemical potentials $\mu_B$ and $\mu_Q$ and then using Eq. (\ref{NAZ}).
This formula is valid for all $(T, \rho, Y_e)$-values, and can be applied for nuclei with all possible mass numbers and charges. We believe that the knowledge about full nuclear ensemble could be quite helpful for accurate calculations of weak reactions with neutrinos and electrons in supernova simulations. 

The mass distributions contain important information about the composition of nuclear matter in the nuclear liquid-gas phase transition (coexistence region). The concept of statistical equilibrium assumes an intensive interaction between the fragments via specific microscopic processes, like absorption and emission of neutrons, which provide equilibration (see, e.g., discussion in Ref.~(\citet{BotvinaMishustin:2010})). In this situation the nuclei may remain hot and have modified properties, such as binding energies, excited states etc, which differ from those in cold isolated nuclei. 
We should note that the data file for mass distributions \textit{"SMSM-EOS-Mass-Distribution-Tables.txt"} is also stored for each calculation. This file is also available in a machine-readable form in the online journal and on web page of SMSM. The file size is 319 MB, it contains 30380  calculations, 31360 lines with space lines and every line has 1003 columns. The first three columns show $T$, $Y_e$ and $\rho/\rho_0$, and columns 4-1003 show mass distributions  for $A=1-1000$. Data are presented for 35 $T$ and 28 $Y_e$ and 31 $\rho/\rho_0$ values with the same order as in Table 1. 

Since we have $35(T)$x$28 (Y_e)=980$ values for 31 $\rho/\rho_0$ values, we grouped 4 density intervals from lower to higher as
\begin{itemize}
\item {Density-1=$\rho/\rho_0=10^{-8}-5.62.10^{-7}$ (8 values),}
\item {Density-2=$\rho/\rho_0=10^{-6}-5.62.10^{-5}$ (8 values),}
\item{Density-3=$\rho/\rho_0=10^{-4}-5.62.10^{-3}$ (8 values),}
\item {Density-4=$\rho/\rho_0=10^{-2}-3.17.10^{-1}$ (7 values). 
}
\end{itemize}
\begin{figure} 
\begin{center}
\includegraphics[width=8cm,height=7cm]{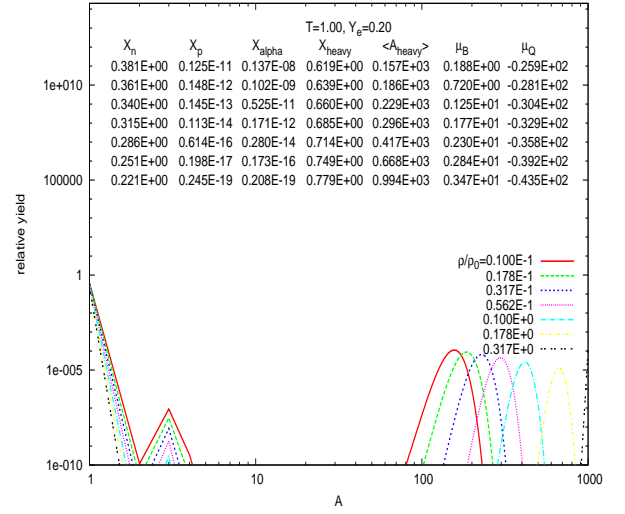}		
\end{center}
\caption{\label{fig_ya1}\small{Mass distributions of nuclear fragments (yields per nucleon:relative yield) produced in stellar matter with 
temperature $T=1$ MeV, electron
fraction $Y_e=0.2$ and at several densities $\rho/\rho_0=10^{-2}-3.17.10^{-1}$. (Color version online.)}}
\end{figure} 
\begin{figure} 
\begin{center}
\includegraphics[width=8cm,height=7cm]{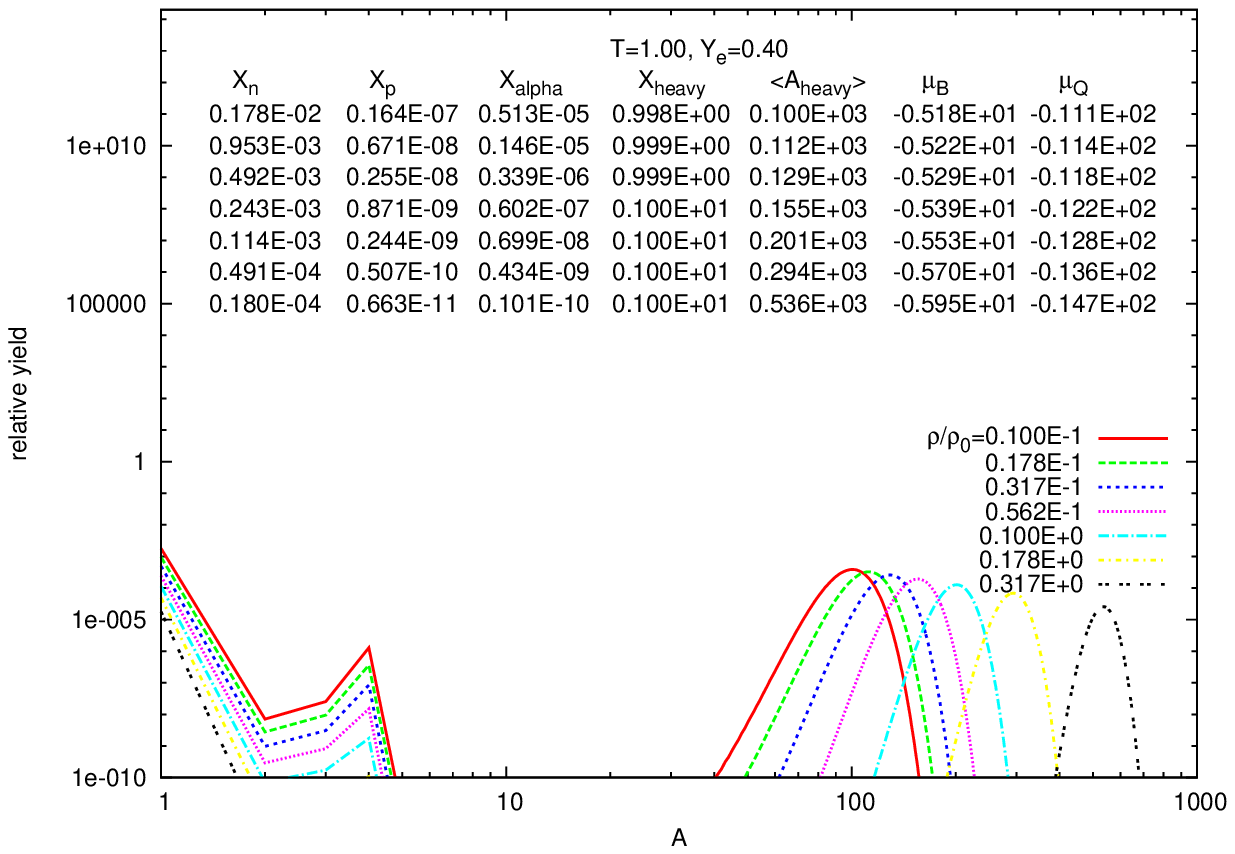}
\end{center}
\caption{\label{fig_ya2}\small{The same as Fig.~\ref{fig_ya1} but at $Y_e=0.4$.}}
\end{figure} 
\begin{figure} 
\begin{center}
\includegraphics[width=8cm,height=7cm]{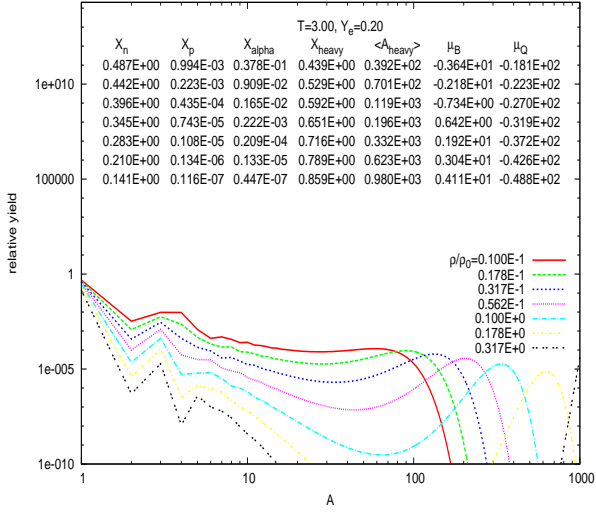}
\end{center}
\caption{\label{fig_ya3}\small{The same as Fig.~\ref{fig_ya1} but at $T=3$ MeV.}}
\end{figure} 
\begin{figure} 
\begin{center}
\includegraphics[width=8cm,height=7cm]{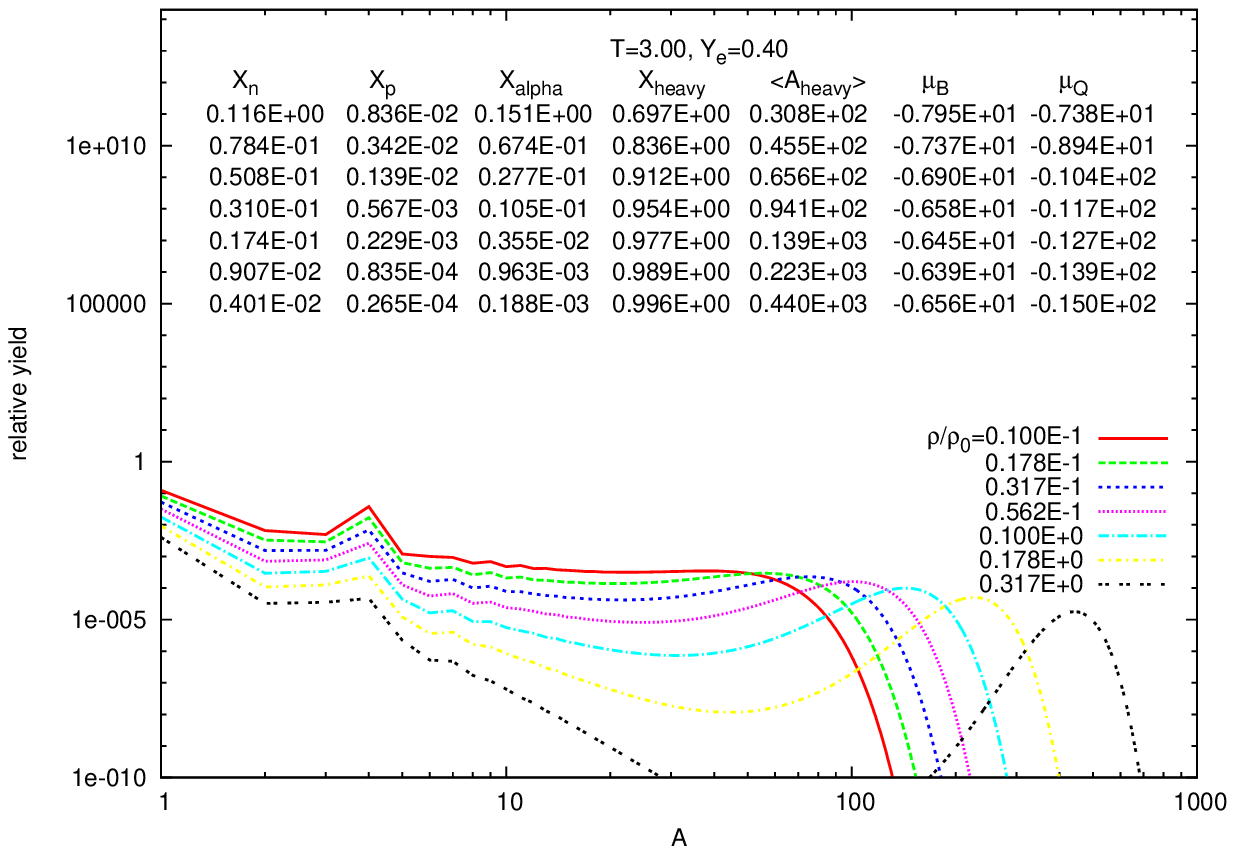}
\end{center}
\caption{\label{fig_ya4}\small{The same as Fig.~\ref{fig_ya3} but at $Y_e=0.4$.}}
\end{figure}  

\begin{figure} 
\begin{center}
\includegraphics[width=8cm,height=14cm]{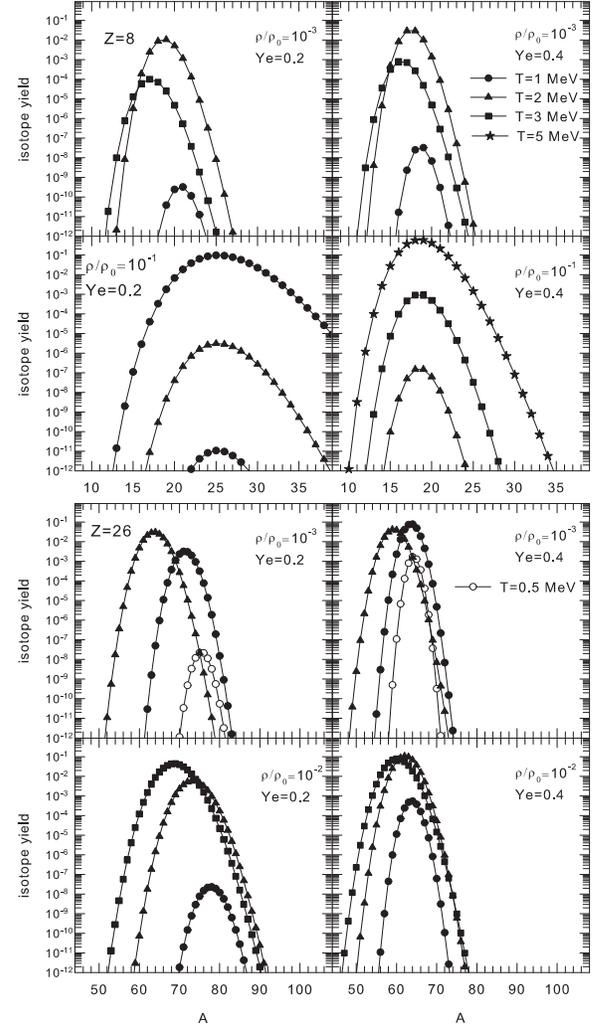}
\end{center}
\caption{\label{fig_iso}\small{Isotopic distributions of $Z=8$ and $Z=26$ fragments produced in matter with 
temperatures $T=1, 2, 3$, and $5$ MeV, electron
fractions $Y_e=0.2$ and $0.4$, densities $\rho/\rho_0=10^{-3}-10^{-1}$.}}
\end{figure} 
\begin{figure} 
\begin{center}
\includegraphics[width=8cm,height=14cm]{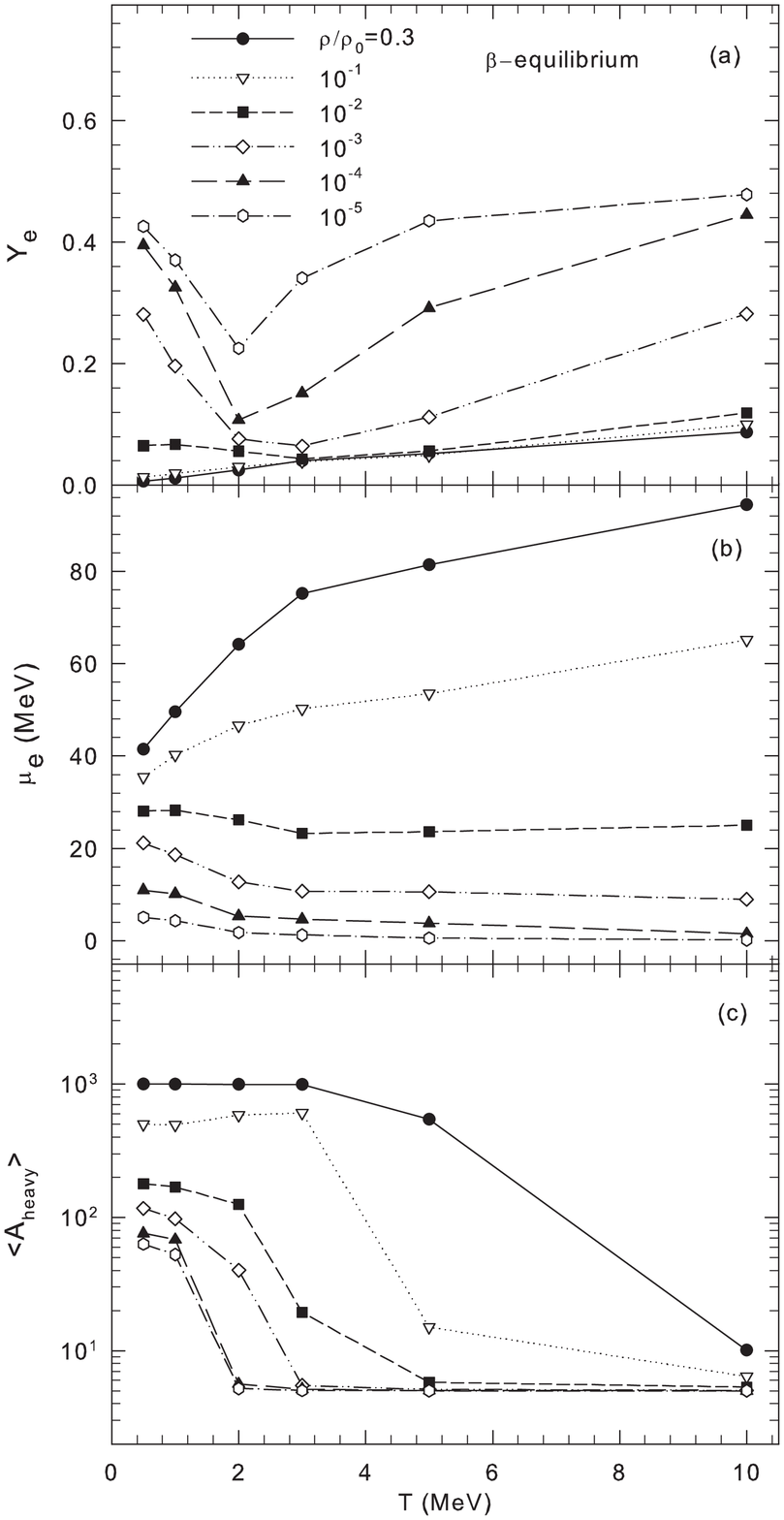}
\end{center}
\caption{\label{fig_ye_beta}\small{Electron fractions, the chemical potential of electrons $\mu_{\rm e}$ and the average value of heavy fragments $<A_{heavy}>$ in $\beta$-equilibrium case versus temperature at densities $\rho/\rho_0=0.3-10^{-5}$.}}
\end{figure} 

For these intervals we have plotted 3920 figures as eps files containing information about mass distributions for each grid point. We have presented example figures for $T=1$, and $3$ MeV, $Y_e=0.2-0.4$ and different densities in Figs. \ref{fig_ya1}, \ref{fig_ya2}, \ref{fig_ya3} and \ref{fig_ya4}. As seen in these figures, we also put $X_n$, $X_p$, $X_{alpha}$, $X_{heavy}$, $<A_{heavy}>$, $\mu_B$ and $\mu_Q$ values from tables with the same order of $\rho/\rho_{0}$ as the control parameters inside of figures. These figures are presented on the SMSM web page with files
\begin{itemize}
\item {\textit{SMSM-Massdist-Figs-Density-1.zip,}}
\item {\textit{SMSM-Massdist-Figs-Density-2.zip,}}
\item {\textit{SMSM-Massdist-Figs-Density-3.zip,}}
\item {\textit{SMSM-Massdist-Figs-Density-4.zip}}
\end{itemize}
Each file contains 980 eps figures. For example, inside \textit{SMSM-Massdist-Figs-Density-4.zip} file, one can find mass distribution figure for $T=1$ MeV, $Y_e=0.20$ and $\rho/\rho_0=10^{-2}-3.17.10^{-1}$ in \textit{T-1-Ye-020-DENSITY-4.eps} file as shown in Fig.~\ref{fig_ya1}.

At very low temperatures the SMSM predicts a Gaussian-like distribution for heavy nuclei as seen in Figs. \ref{fig_ya1} and \ref{fig_ya2}. In this case an approximation of a single heavy nucleus adopted in the EOS by \citet{LattimerSwesty:1991} and \citet{Shen} may work reasonably well for calculations of thermodynamical characteristics of matter. However, already at $T \geq 1$ MeV the gap between the Gaussian peak and light clusters and nucleons is essentially filled by nuclei of intermediate masses leading to characteristic U-shaped distributions. The examples of such distributions are shown in Figs. \ref{fig_ya3} and \ref{fig_ya4}. They give a typical evolution of mass distributions of nuclei in the coexistence region of the liquid gas phase transition. Previously it was demonstrated for disintegration of finite nuclei (\citet{Bondorf}). These distributions continuously transform with decreasing density as well as with increasing temperatures into exponential fall-off of fragment yields as functions of $A$.

Using SMSM EOS tables and Eq.~(\ref{NAZ}) it is also possible to obtain isotopic distributions of stellar matter. Examples of isotopic distributions for $Z=8$ and $Z=26$ are shown in Fig.~\ref{fig_iso}. As seen from the figure, the SMSM predicts Gaussian type distributions, that is a consequence of the liquid-drop description of fragments. 
Isotopic yields help to understand the trends observed for summed quantities like mass yields or mass fractions.
They are also needed to calculate average $<Z>$ values which are listed in the tables. This information is important for realistic calculations of the weak reactions with electrons and neutrinos. 

Recently, in Ref.~(\citet{nihal2013}) we have compared mass and isotope distributions of the SMSM (\citet{BotvinaMishustin:2010}) with predictions of two other models by  \citet{HempelSchaffnerBielich:2010} and \citet{furusawa11}, under conditions expected during the collapse of massive stars and supernova explosions.  Presently, we have found significant differences between mass distributions predicted by these three models, which use different assumptions on properties of hot fragments in a dense environment, especially at low electron fractions, low temperatures and high densities. We note that the differences are also present in the behaviour of nuclear pressure, see Fig.~22 in Ref.~(\citet{nihal2013}). 

Indeed, the main assumption of the SMSM is that most interaction effects are
included into the internal binding energy of the fragments. Apart of that, we
also take into account the Coulomb energy and excluded volume effects. As one
can see from the tables, the mass fraction of single nucleons is rather small
(less than 20 per cent) at baryon densities around $0.1 \rho_0$ and electron fractions
around 0.2.  Their actual density is therefore only $0.02 \rho_0$. The interaction
effects are surely negligible at such a low density, see e.g. (\citet{Talahmeh&Jaqaman:2013}). 
At higher densities and lower electron fractions the share of unbound nucleons becomes even
larger and their interaction may become noticeable. But even at baryon
densities around $0.3 \rho_0$ and temperatures $T\sim 1$ MeV most nucleons are bound in clusters 
and most interaction effects come from the nuclear binding energy. Exactly in this domain of
parameters different models predict very different nuclear ensembles, see (\citet{nihal2013}). 
Certainly, improving the EOS calculations in this region is an important issue for future studies. 
Recently \citet{furusawa13} have made an attempt to improve the description of nucleons and 
light nuclei in supernova matter.

\subsection{Nuclear composition under condition of $\beta$-equilibrium}

An important special case when full $\beta$-equilibrium is reached in stellar matter corresponds to the condition $\mu_e=\mu_n-\mu_p$. It is expected to be fulfilled in certain situations like slow collapse of a massive star, late stages of a supernova explosion or in crusts of neutron stars. On the other hand, the $\beta$-equilibrium can be a useful physical limit for theoretical estimates of the nuclear composition without full knowledge of weak reactions. The SMSM allows for this kind of calculations (\citet{BotvinaMishustin:2004,BotvinaMishustin:2010}) and the corresponding Tables of matter properties at baryon densities $\rho= (10^{-8}-0.32)\rho_0$, and temperatures $T=0.2-25$ MeV are under construction. At given $\rho$ and $T$ the electron fraction $Y_e$ can be evaluated iteratively to fulfil the above-mentioned condition. 

In Fig.~\ref{fig_ye_beta} we show the calculated electron fraction, the chemical potential of electrons $\mu_{\rm e}$ and the average mass number of heavy nuclei $<A_{heavy}>$ $(A>4)$ under $\beta$-equilibrium condition as function of temperature for various densities. Fig.~8 gives a valuable information about the composition of stellar matter under condition of $\beta$-equilibrium. In Fig. 8a one can see an interesting behaviour, i.e. a non-monotonic change of $Y_e$ with increasing temperature, which has a simple explanation. As follows from Fig. 8b, electrons have larger chemical potentials at higher densities ($\rho=0.1-0.3\rho_0$) (Fig.8b), when heavy nuclei (Fig.8c) can survive even at high temperatures $T=6-8$ MeV. That is why the fraction of electrons $Y_e$ increases slowly with temperature. On the other hand, at lower  densities ($\rho \leqslant 0.01\rho_0$) the heavy nuclei can exist only at low temperatures and they disintegrate into free nucleons and light clusters at temperatures $T>3-5$ MeV. However, in Fig. 8a, one can see that  $Y_e$ exhibits a minimum at $T\approx 2$ MeV for $10^{-5}\div 10^{-4}\rho_0$ and $T =3\div 4$ MeV for $(10^{-3}\div 10^{-2}\rho_0$. As follows from Fig. 8c, these temperatures correspond to the transition from heavy nuclei to free nucleons and light clusters. At higher densities, ($\rho= (0.3\div 0.1)\rho_0$), when heavy nuclei are present in the ensemble even at higher temperatures, the electron fraction grows slowly with the temperature, reaching ($10\div 12$)\% at $T=10$ Mev. These results reveal an interesting correlation between the nuclear composition and electron fraction in $\beta$-equilibrated stellar matter.

\section{Conclusions}

We have presented the SMSM EOS tables which give comprehensive information on physical properties of stellar matter at subnuclear densities, temperatures $T<25$ MeV and electron fractions between 0.02 and 0.56. The SMSM EOS can be used for hydrodynamical simulations of massive star collapse and supernova explosion processes. We consider the whole ensemble of nuclear species, without any artificial constrains on their masses and charges, as well as thermodynamical characteristics of matter, such as temperature and chemical potentials. The nuclear mass and isotope distributions are needed for realistic calculations of electron capture and neutrino scattering processes in supernova environments. 
The SMSM is directly linked to nuclear multifragmentation reactions which allow to study in laboratory the properties of nuclei embedded in hot and dense surrounding. The present tabulated calculations are performed at the standard 
parameters of nuclei. However, as was demonstrated previously 
(see \citet{BotvinaMishustin:2004,BotvinaMishustin:2010}), modifications of these 
parameters, in particular, the symmetry energy of nuclei extracted 
from multifragmentation reactions, can essentially change the nuclear composition and influence the rate of weak reactions.  Our calculations clearly demonstrate that the nuclear composition is very sensitive to temperature, density, and electron fraction of stellar matter. A special case of $\beta$-equilibrated nuclear matter is considered too. In this case  we have found non-monotonous behavior of the electron fraction as a function of temperature at $\rho<0.2\rho_0$. This effect is connected with the rearrangement of the fragment mass distribution from U-shape to exponential fall-off. These results can be used for modelling the outer layers of proto-neutron stars and crusts of neutron stars. We believe that the comparison of our results with predictions of other models, as was done in \citet{nihal2013}, will help to better understand properties of hot and dense stellar matter.

\section{Acknowledgement}

N.B. and A.S.B. are grateful to Frankfurt Institute for Advanced Studies (FIAS) for support and hospitality. This work is supported by the Helmholtz International Center for FAIR within the LOEWE program. N.B. thanks Selcuk University-Scientific Research Projects (BAP) for partial support. N.B. acknowledges that a part of the numerical calculations was carried out on the computers at Physics Department of Selcuk University. N.B. is grateful to R.Ogul for encouragement, and to A.E. Kavruk and T. Ozturk for the help in computing. A.S.B. acknowledges the support by the Research Infrastructure Integrating Activity 'Study of Strongly Interacting Matter' HadronPhysics3 under the 7th Framework Programme of EU. I.N.M. acknowledges support from the grant NSH-932.2014.2 (Russia).


\begin{thebibliography}{99}

\bibitem[{Arcones} et~al. (2008)] {Arcones:etal:2008} 
{Arcones},~A., {Mart{\'{\i}}nez-Pinedo},~G., {O'Connor},~E. et~al.
\newblock 2008,
\newblock \emph{\prc}, 78, 015806

\bibitem[{Avdeyev} et~al. (2002)] {FASA} 
{Avdeyev},~S.~P., {Karnaukhov},~V.~A., {Petrov},~L.~A. et~al. 
\newblock 2002,
\newblock \emph{Nucl. Phys. A}, 709, 392

\bibitem[{Baym}, {Pethick}, \& {Sutherland} (1971)] {BPS} 
{Baym},~G., {Pethick},~C., \& {Sutherland},~P.
\newblock 1971,
\newblock \emph{\apj}, 170, 299

\bibitem[{Bellaize} et~al. (2002)] {INDRA} 
{Bellaize}, N., {Lopez},~O., {Wieleczko},~J.~P. et al.
\newblock 2002,
\newblock \emph{Nucl. Phys. A}, 709, 367


\bibitem[{Bondorf} et~al. (1985)] {Bondorf85} 
{Bondorf},~J.~P., {Donangelo},~R., {Mishustin},~I.~N., {Pethick},~C.~J., {Schulz},~H. \& {Sneppen},~K.
\newblock 1985,
\newblock \emph{Nucl. Phys. A}, 443, 321


\bibitem[{Bondorf} et~al. (1995)] {Bondorf} 
{Bondorf},~J.~P., {Botvina},~A.~S., {Iljinov},~A.~S., {Mishustin},~I.~N., \& {Sneppen},~K.
\newblock 1995,
\newblock \emph{\physrep}, 257, 133

\bibitem[{Botvina}, {Iljinov}, \& {Mishustin} (1985)] {Botvina85} 
{Botvina},~A.~S., {Iljinov},~A.~S., \& {Mishustin},~I.~N.
\newblock 1985,
\newblock \emph{Sov. J. Nucl. Phys.}, 42, 712

\bibitem[{Botvina}, {Iljinov}, \& {Mishustin} (1990)] {Botvina90} 
{Botvina},~A.~S., {Iljinov},~A.~S., \& {Mishustin},~I.~N.
\newblock 1990,
\newblock \emph{Nucl. Phys. A}, 507, 649

\bibitem[{Botvina} et~al. (1995)] {ALADIN} 
{Botvina},~A.~S., {Mishustin},~I.~N., {Begemann-Blaich},~M. et~al.
\newblock 1995,
\newblock \emph{Nucl. Phys. A}, 584, 737


\bibitem[{Botvina}, {Lozhkin}, \& {Trautmann} (2002)] {traut} 
{Botvina},~A.~S., {Lozhkin},~O.~V., \& {Trautmann},~W. 
\newblock 2002,
\newblock \emph{Phys. Rev. C}, 65, 044610

\bibitem[{Botvina} \& {Mishustin} (2004)] {BotvinaMishustin:2004}
{Botvina},~A.~S. \& {Mishustin},~I.~N.
\newblock 2004,
\newblock \emph{Phys. Lett. B}, 584, 233

\bibitem[{Botvina} et~al. (2006)] {Botvina06} 
{Botvina},~A.~S., {Buyukcizmeci},~N., {Erdogan},~M. et~al.
\newblock 2006,
\newblock \emph{\prc}, 74, 044609

\bibitem[{Botvina} \& {Mishustin} (2010)] {BotvinaMishustin:2010}
{Botvina},~A.~S. \& {Mishustin},~I.~N.
\newblock 2010,
\newblock \emph{Nucl. Phys. A}, 843, 98

\bibitem[{Sagun} et al. (2014)] {Sagun2014} 
{Sagun},~V.~V., {Ivanytskyi}, ~A.~I, {Bugaev},~K.~A.\& {Mishustin},~I.~N.
\newblock 2014,
\newblock \emph{Nucl. Phys. A}, 924, 24


\bibitem[{Bugaev}, {Gorenstein}, \& {Mishustin} (2001)] {Bugaev} 
{Bugaev},~K.~A., {Gorenstein},~M.~I., \& {Mishustin},~I.~N.
\newblock 2001,
\newblock \emph{Phys. Lett. B}, 498, 144

\bibitem[{Buyukcizmeci} et~al. (2008)] {bulk} 
{Buyukcizmeci},~N., {Botvina}~A.~S., {Mishustin}~I.~N., \& {Ogul},~R.
\newblock 2008,
\newblock \emph{Phys. Rev. C}, 77, 034608

\bibitem[{Buyukcizmeci} et~al. (2013)] {nihal2013} 
{Buyukcizmeci},~N., {Botvina},~A.~S., {Mishustin}~I.~N. et~al.
\newblock 2013,
\newblock \emph{Nucl. Phys. A}, 907, 13

\bibitem[{D'Agostino} et~al. (1996)] {MSU} 
{D'Agostino},~M., {Botvina},~A.~S., {Milazzo},~P.~M. et~al.
\newblock 1996,
\newblock \emph{Phys. Lett. B}, 371, 175

\bibitem[{D'Agostino} et~al. (1999)] {Dag} 
{D'Agostino},~M., {Botvina},~A.~S., {Bruno}, M. et~al.
\newblock 1999,
\newblock \emph{Nucl. Phys. A}, 650, 329

\bibitem[{Eur. Phys. J. A, 30} (2006)] {EPJA}
Dynamics and Thermodynamics with Nuclear Degrees of Freedom, 
ed. by {Chomaz},~Ph., {Gulminelli},~ F., {Trautmann},~ W., \& {Yennello},~S.~J. (Springer, Berlin/Heidelberg/New York)
\newblock 2006,
\newblock \emph{Eur. Phys. J. A}, 30

\bibitem[{Furusawa} et~al. (2011)] {furusawa11} 
{Furusawa},~S., {Yamada},~S., {Sumiyoshi},~K., \& {Suzuki},~H.
\newblock 2011,
\newblock \emph{\apj}, 738, 178


\bibitem[{Furusawa} et~al. (2013)] {furusawa13} 
{Furusawa},~S., {Sumiyoshi},~K., {Yamada},~S., \& {Suzuki},~H.
\newblock 2013,
\newblock \emph{\apj}, 772, 95

\bibitem[{Gross} (1982)]{Gross82}
{Gross},~D.~H.~E.
\newblock 1982,
\newblock \emph{Z. Phys. A}, 309, 41


\bibitem[{Gross} (1990)]{Gross}
{Gross},~D.~H.~E.
\newblock 1990,
\newblock \emph{Rep. Prog. Phys.}, 53, 605

\bibitem[{Hauger} et~al. (2000)] {Hauger} 
{Hauger},~J.~A., {Srivastava},~B.~K., {Albergo},~S. et~al. 
\newblock 2000,
\newblock \emph{Phys. Rev. C}, 62, 024616

\bibitem[{Hempel} \& {Schaffner-Bielich} (2010)] {HempelSchaffnerBielich:2010}
{Hempel},~M. \& {Schaffner-Bielich},~J.
\newblock 2010,
\newblock \emph{Nucl. Phys. A}, 837, 210

\bibitem[{Hempel} et~al. (2011)] {Hempel11b} 
{Hempel},~M., {Schaffner-Bielich},~J., {Typel},~S., \& {R{\"o}pke},~G.
\newblock 2011,
\newblock \emph{Phys. Rev. C}, 84, 055804

\bibitem[{Hempel} et~al. (2012)] {Hempel12} 
{Hempel},~M., {Fisher},~T., {Schaffner-Bielich},~J., \& {Liebend{\"o}rfer},~M.
\newblock 2012,
\newblock \emph{Astrophys. J.}, 70, 748

\bibitem[{Hudan} et~al. (2009)] {Hudan} 
{Hudan},~S., {McIntosh},~A.~B., {Black},~J. et~al.
\newblock 2009,
\newblock \emph{\prc}, 80, 064611

\bibitem[{Iglio} et~al. (2006)] {Iglio} 
{Iglio},~J., {Shetty},~D.~V., {Yennello},~S.~J. et~al. 
\newblock 2006,
\newblock \emph{\prc}, 74, 024605

\bibitem[{Ishizuka}, {Ohnishi}, \& {Sumiyoshi} (2003)] {Japan} 
{Ishizuka},~C., {Ohnishi},~A., \& {Sumiyoshi},~K.
\newblock 2003,
\newblock \emph{Nucl. Phys. A}, 723, 517

\bibitem[{Lamb} et~al. (1981)] {Lamb:1981} 
Lamb,~D.~Q., Lattimer,~J.~M., \& Pethick,~C.~J.
\newblock 1981,
\newblock \emph{Nucl. Phys. A}, 360, 459

\bibitem[{Lattimer} et~al. (1985)] {Lattimer:etal:1985} 
Lattimer,~J.~M., Pethick,~ C.~J., Ravenhall,~ D.~G., \& Lamb,~D.~Q. 
\newblock 1985,
\newblock \emph{Nucl. Phys. A}, 432, 646

\bibitem[{Lattimer} \& {Swesty} (1991)] {LattimerSwesty:1991}
{Lattimer},~J.~M. \& {Swesty},~D.~F.
\newblock 1991,
\newblock \emph{Nucl. Phys. A}, 535, 331


\bibitem[{Newton} \& {Stone} (2009)] {Newton} 
{Newton},~J.~R., {Stone},~W.~G.
\newblock 2009,
\newblock \emph{\prc}, 79, 055801


\bibitem[{Ogul} et~al. (2011)] {Ogul} 
{Ogul},~R., {Botvina},~A.~S., {Atav}~U. et~al.
\newblock 2011,
\newblock \emph{\prc}, 83, 024608

\bibitem[{Pienkowski} et~al. (2002)] {Pienkowski} 
{Pienkowski},~L., {Kwiatkowski},~K., {Lefort},~T. et~al.
\newblock 2002,
\newblock \emph{\prc}, 65, 064606

\bibitem[{Prakash} et~al. (1997)] {Prakash} 
{Prakash},~M., {Bombaci},~I., {Prakash},~M. et~al. 
\newblock 1997,
\newblock \emph{Phys. Rep.}, 280, 1

\bibitem[{Randrup and Koonin (1981)}] {Randrup81}
{Randrup},~J. \& {Koonin},~S.E. 
\newblock 1981,
\newblock \emph{Nucl. Phys. A}, 356, 223

\bibitem[{Rodionov} et~al. (2002)] {Rodionov} 
{Rodionov},~V., {Avdeyev},~S.~P., {Karnaukhov},~V.~A. et~al. 
\newblock 2002,
\newblock \emph{Nucl. Phys. A}, 700, 457

\bibitem [{Sagun} et~al. (2013)] {Sagun} 
{Sagun},~V.V., {Ivanytskyi},~A.I., {Bugaev},~K.A., {Mishustin},~I.N.
\newblock 2013,
\newblock \emph{Nucl. Phys. A}, 924, 24 

\bibitem[{Scharenberg} et~al. (2001)] {EOS} 
{Scharenberg},~R.~P., {Srivastava},~B.~K., {Albergo},~S. et~al.
\newblock 2001,
\newblock \emph{\prc}, 64, 054602

\bibitem[{Shen} et~al. (1998)] {Shen} 
{Shen},~H., {Toki},~H., {Oyamatsu},~K., \& {Sumiyoshi},~K.
\newblock 1998,
\newblock \emph{Nucl. Phys. A}, 637, 435

\bibitem[{Souliotis} et~al. (2007)] {Souliotis} 
{Souliotis},~G.~A., {Botvina},~A.~S., {Shetty},~D.~V. et~al.
\newblock 2007,
\newblock \emph{\prc}, 75, 011601

\bibitem[{Sugahara} \& {Toki} (1994)] {Suga94} 
{Sugahara},~Y. \& {Toki},~H.
\newblock 1994,
\newblock \emph{Nucl. Phys. A}, 637, 557

\bibitem[{Sumiyoshi} et~al. (2005)] {Sumi} 
{Sumiyoshi},~K., {Yamada},~S., {Suzuki},~H. et~al.
\newblock 2005,
\newblock \emph{\apj}, 629, 922

\bibitem[{Sumiyoshi} \& {R{\"o}pke} (2008)] {SumiyoshiRopke:2008}
{Sumiyoshi},~K. \& {R{\"o}pke},~G. 
\newblock 2008,
\newblock \emph{\prc}, 77, 055804

\bibitem[{Talahmeh} \& {Jaqaman} (2013)] {Talahmeh&Jaqaman:2013}
{Talahmeh}, S. \& {Jaqaman}, H.R.
\newblock 2013,
\newblock \emph{J. Phys. G: Nucl. Part. Phys.}, 40,  015103

\bibitem[{Typel} et~al. (2010)] {Typel:etal:2010} 
{Typel},~S., {R{\"o}pke},~G., {Kl{\"a}hn},~T., {Blaschke},~D., \& {Wolter},~H.~H.
\newblock 2010,
\newblock \emph{\prc}, 81, 015803

\bibitem[{Viola} et~al. (2001)] {Viola} 
{Viola},~V., {Lefort},~T., {Beaulieu},~L. et~al.
\newblock 2001,
\newblock \emph{Nucl. Phys. A}, 681, 267

\bibitem[{Wang} et~al. (1999)] {Wang} 
{Wang},~G., {Kwiatkowski},~K., {Bracken},~D.~S. et~al.
\newblock 1999,
\newblock \emph{\prc}, 60, 014603

  
\end{thebibliography}
\end{document}